# Inequality in Education: A Comparison of Australian Indigenous and Nonindigenous Populations

David Gunawan*
*University of Wollongong*

William Griffiths
*University of Melbourne*

Duangkamon Chotikapanich
*Monash University*

26 August, 2021

**Abstract**

Educational achievement distributions for Australian indigenous and nonindigenous populations in the years 2001, 2006, 2014 and 2017 are considered. Bayesian inference is used to analyse how these ordinal categorical distributions have changed over time and to compare indigenous and nonindigenous distributions. Both the level of educational achievement and inequality in educational achievement are considered. To compare changes in levels over time, as well as inequality between the two populations, first order stochastic dominance and an index of educational poverty are used. To examine changes in inequality over time, two inequality indices and generalised Lorenz dominance are considered. Results are presented in terms of posterior densities for the indices and posterior probabilities for dominance for the dominance comparisons. We find some evidence of improvement over time, especially in the lower parts of the indigenous distribution and that inequality has significantly increased from 2001 to 2017.

*Corresponding author
David Gunawan
School of Mathematics and Applied Statistics
University of Wollongong, NSW 2522
Australia
dgunawan@uow.edu.au

# 1. Introduction

Improving the general level of education and reducing inequality in education are worthy objectives of public policy. A higher level of education increases one's capacity to generate income and improves one's prospects for a more fulfilling and happy life. The existence of educational inequalities, where there are wide disparities between the highly educated and those with lower levels of education, or where one subgroup of society has lower levels of education than the remainder of the population, can lead to distrust, discrimination, resentment, and feelings of helplessness. To improve the level of welfare of its indigenous population (Aboriginal and Torres Strait Islanders), and to reduce inequality between its indigenous and nonindigenous populations, the Australian government has introduced a program called "Closing the Gap" (www.closingthegap.gov.au). This program has 17 targets related to health, education, employment, culture and general well-being. We focus on education. Statistical methodology for assessing whether improvements in the level of education and reductions in educational inequality have been realised over the period 2001-2017 is introduced. This methodology is applied to data extracted from the Household and Labour Dynamics in Australia (HILDA) survey for the years 2001, 2006, 2014 and 2017.[1]

Various indicators used to measure education status have been described and reviewed by Thomas et al. (2001). The impacts of socio-economic background and the availability of resources on educational inequality have received considerable attention in the literature. For access to this literature, and a discussion of the many dimensions of inequality, see, for example, Angelico (2020), Dean (2019) and Perry (2018). For measuring education status, we focus on the highest education level achieved by an individual, measured on the ordinal categorical scale displayed in Table 1. The numbers of individuals in each of these categories for each year, and for the indigenous and nonindigenous subgroups, are taken from the HILDA data. They represent samples from ordinal categorical distributions, used to make inferences about their corresponding populations. These inferences take the form of comparing education status at different points in time and comparing the education status of indigenous and nonindigenous subgroups.

---

[1] The HILDA survey is a national representative longitudinal survey which began in Australia in 2001; it is designed managed and maintained by the Melbourne Institute of Applied Economic and Social Research, University of Melbourne (Watson and Wooden, 2012).

The discrete nature of the distributions prevents the use of traditional tools such as the Lorenz curve and the Gini coefficient that are used for measuring inequality in continuous income distributions. Alternatives that have been suggested for ordinal data, and that we use in this investigation, are the indices proposed by Cowell and Flachaire (2020) and Jenkins (2020) who illustrate their proposals with applications to data on life satisfaction and health. We denote them by *CF* and *J*, respectively. Other approaches that have appeared in the literature are those of Apouey et al. (2020) and Anderson et al. (2020). We also consider the proportion of individuals who fail to reach a particular standard of education – those in categories 1 and 2 in our distribution. In the Australian education system, category 1 represents those who do not complete the final year of high school; category 2 contains those who complete the final year of high school but do not proceed to tertiary education. Borrowing terminology from poverty analysis, we refer to the proportion of individuals in categories 1 and 2 as the headcount index, *H*. This index is particularly relevant for the closing-the-gap program where one of the 2031 targets is to decrease *H* to 0.3 for those in the indigenous population aged 25 to 34. In addition to comparing educational distributions using the *CF*, *J* and *H* indices, we also employ dominance concepts. The *CF*, *J* and *H* indices are useful for providing a complete ordering of the distributions, but they may not reveal some important features uncovered by comparing complete distributions. An alternative approach that considers the full support of the distributions is to use the concept of stochastic dominance. We consider both first order stochastic dominance (FSD), as described by Allison and Foster (2004), and generalised Lorenz dominance (GLD), as proposed by Jenkins (2020).

Our objective is to compare distributions over time and between indigenous and non-indigenous populations to shed light on the following four questions.

1. Has the level of education of the indigenous population improved over time? For this purpose, we consider FSD and *H*.

2. Has inequality between the indigenous and non-indigenous populations declined? The tools FSD and *H* are also used to answer this question.

3. Has inequality within the indigenous population changed? For this case, we use GLD, *CF* and *J*.

4. How does inequality within the indigenous population compare with inequality within the non-indigenous population? Here, again, GLD, *CF* and *J* are relevant.

Our approach is Bayesian. We use the algorithm described in Gunawan et al. (2018) for estimating the ordered categorical distributions' probabilities, combined with a Bayesian bootstrap algorithm (Gunawan, et al., 2020) that incorporates the sampling weights provided by the HILDA data. Results for *CF*, *J* and *H* are presented in terms of their posterior densities. Results for FSD and GLD are presented in terms of posterior probabilities of dominance. An advantage of the Bayesian approach is that three dominance probabilities can be reported. Given two education distributions *X* and *Y*, we can report the probability that *X* dominates *Y*, the probability that *Y* dominates *X*, and the probability that neither distribution is dominant. The reporting of dominance probabilities is a valuable way of summarising the results of the dominance assessments. Whether a probability is sufficiently high for a policymaker to take a particular action is the policymakers decision. The posterior densities for *CF*, *J* and *H* can also be used to make probability statements about their values.

In Section 2, the dominance criteria and inequality measures are described. The Bayesian approach for making inferences about dominance criteria and inequality measures is discussed in Section 3. The results are presented in Section 4. Section 5 concludes. The paper also has an online supplement with details of the estimation procedure.

## 2 Dominance and Inequality Measures for Ordinal Data

In an ordered categorical distribution, each of a sample of $N$ individuals is assigned an ordered, but otherwise arbitrary, numerical label. For our application where there are $K=7$ categories, we use the labels $k=1,2,\ldots,7$. Let the population proportion of individual in category $k$ be given by $p(k)$, with $0 \leq p(k) \leq 1$, and $\sum_{k=1}^{K} p(k) = 1$. The cumulative distribution function is denoted by $F(k) = \sum_{j=1}^{k} p(j)$.

The FSD criterion proposed by Allison and Foster (2004) is concerned with whether one distribution represents a higher level of health relative to another. In our context it can be used to compare levels of education. Given any two discrete education variables, $X$ and $Y$, with the same total number of categories $K$, we say $X$ first order stochastic dominates $Y$, written as, $X >_{FSD} Y$, if and only if

$$F_X(k) \leq F_Y(k) \text{ for all } k \text{ and } F_X(k) < F_Y(k) \text{ for some } k \qquad (1)$$

for $k = 1, 2, \ldots, K-1$. The distribution of $X$ has a lower percentage of its population in the lower $k$ categories than distribution $Y$. Hence, there is a higher level of education in $X$ than in $Y$.

If we are particularly interested in improving the education status of individuals in the lower part of the distribution, then two relevant criteria, less strict than the FSD criterion, are the *H* index and a restricted form of dominance. If more individuals from distribution *X*, than from distribution *Y*, are completing some form of tertiary education, then $H_X < H_Y$ where $H_X = p_X(1) + p_X(2)$ and $H_Y = p_Y(1) + p_Y(2)$. We have a restricted form of FSD if $H_X < H_Y$ and $p_X(1) < p_Y(1)$; more individuals in *X* are completing year 12 or higher, *and* more individuals in *X* are completing a tertiary qualification.

The remaining measures, *CF*, *J* and *GLD* are concerned with inequality in education. Using a generalized entropy index, Cowell and Flachaire (2017) consider inequality from the standpoint of an individual's perception of their status in the distribution. Indices based on "downward-looking" and "upward-looking" assessment of status and which satisfy several desirable axioms are developed. We consider the downward-looking index which is an aggregate summary of the difference between each person downward-looking status and 1 (the maximum value of cumulative distribution function). Their one-parameter family of inequality indices, $CF(\alpha)$, with $0 \leq \alpha < 1$ is

$$CF(\alpha) = \begin{cases} \frac{1}{\alpha(\alpha-1)} \left[ \sum_{k=1}^{K} p(k) \left[ \sum_{j=1}^{k} p(j) \right]^{\alpha} - 1 \right], & \text{if } \alpha \neq 0, 1 \\ -\sum_{k=1}^{K} p_k \log \left[ \sum_{j=1}^{k} p(j) \right] & \text{if } \alpha = 0 \end{cases} \qquad (2)$$

The smaller the value of parameter $\alpha$, the greater the weight put on small status values relative to that on high status values.

An ordering of distributions based on the GLD concept introduced by Jenkins (2020) implies the same ranking of distributions according to all members of the class of *CF* indices. Jenkins' generalized Lorenz curve $GL(u)$ consists of a series of linear segments plotted against the population proportion $0 \leq u \leq 1$. Using $F(k)$ as the education status of an individual in category $k$, the first linear segment of $GL(u)$ is $GL(u,1) = uF(1)$. The second linear segment, between the two points $u = F(1)$ and $u = F(2)$, is given by $GL(u,2) = p(1)F(1) + [u - F(1)]F(2)$, and the $k$-th linear segment, between $u = F(k-1)$ and $u = F(k)$, is

$$GL(u,k) = \sum_{i=1}^{k} p(i-1)F(i-1) + [u - F(k-1)]F(k) \qquad (3)$$

Combining all segments, we have

$$GL(u) = \sum_{k=1}^{K} \left\{ \left[ I\left(u \geq F(k-1)\right) \right] p(k-1)F(k-1) + \left[ I\left(F(k-1) \leq u \leq F(k)\right) \right] [u - F(k-1)]F(k) \right\} \qquad (4)$$

where $I(\cdot)$ is an indicator function equal to one if its argument is true, and zero otherwise. If all individuals have the same education status $[p(k) = F(k) = 1$ for any $k]$, the $GL$ curve is a 45-degree line through the origin and point (1,1). We have complete equality. If there is inequality in the distribution, the $GL$ curve is below this 45-degree line with its endpoint at $u = F(k)$ given by $GL(1) = \sum_{k=1}^{K} p(k)F(k)$. We say $X$ generalized Lorenz dominates $Y$, written as $X >_{GLD} Y$, if and only if

$$GL_X(u) \geq GL_Y(u) \text{ for all } 0 \leq u \leq 1 \text{ and } GL_X(u) > GL_Y(u) \text{ for some } u. \qquad (5)$$

The distribution of $X$ has lower inequality than the distribution of $Y$. As examples, in Figure 1 we use the posterior means for the $p(k)$ to plot estimated $GL$ curves for the indigenous population for the years 2001 and 2017. These plots suggest the 2001 distribution dominates that for 2017. In Section 4 we discover that there is indeed strong evidence that inequality has increased from 2001 to 2017.

The $J$ inequality index developed by Jenkins (2020) is equal to twice the area between the 45° line and the $GL$ curve. As with the $CF$ index, a ranking of distributions based on the $J$ index is implied by a $GLD$ ranking. It is given by

$$J = 1 - \sum_{k=0}^{K} p(k+1)\left[GL\big(F(k)\big) + GL\big(F(k+1)\big)\right] \qquad (6)$$

The minimum value of $J$ is zero when there is perfect equality.

## 3 Bayesian Inference

The first step towards finding Bayesian estimates for the inequality measures $CF$ and $J$, and the headcount index $H$, and finding posterior probabilities for FSD and GLD, is to use the sample observations to estimate the population proportions $\boldsymbol{p}' = \big(p(1), p(2), \ldots, p(K)\big)$. We do so by combining the sampling algorithm described in Gunawan et al. (2018) with a Bayesian bootstrap method designed to incorporate sampling weights into the estimation algorithm (Gunawan et al. 2020). A full description of the algorithm is provided in Appendix S1.

The estimation algorithm produces draws $\boldsymbol{p}^{(m)}, m=1,2,\ldots,M$ from its posterior density. These draws are then used to compute corresponding values for $CF, J$ and $H$ from their posterior densities. The means of these values represent point estimates for the $CF, J$ and $H$ indices. The standard deviations reflect the reliability of the point estimates. The draws on the indices can also be used to plot kernel density estimates of their posterior densities.

For the inequality given in Equation (1), we estimate the posterior probability that $X$ first order dominates $Y$, $\Pr[X >_{FSD} Y]$, as the proportion of $M$ draws for which $F_X^{(m)}(k) - F_Y^{(m)}(k) \leq 0$, for all $k=1,2,\ldots,K-1$. For the GLD criteria in Equation (5), we compare the GLDs for $X$ and $Y$ for 99 values of $u$, from 0.01 to 0.99 at intervals 0.01 and estimate the posterior probability $\Pr[X >_{GLD} Y]$ as the proportion of $M$ draws for which $GL_X^{(m)}(u) - GL_Y^{(m)}(u) \geq 0$, for all grid points $u$. Counting the proportions of draws with the inequalities reversed, yields estimates for probabilities of $Y$ dominating $X$. The posterior probability that neither distribution dominates is, for example, $1 - \Pr[X >_{FSD} Y] - \Pr[Y >_{FSD} X]$. A by-product of the estimation procedure is plots of probability curves that give the probability of dominance at a given category $k$ for FSD, and at a given population proportion $u$ for the GLD criterion. The probability of dominance over any range of categories or any range of $u$ will be no greater than the minimum value of the probability curve within that range. This makes the probability curve a valuable device for finding the population proportions or categories which have the greatest impact on the probability of dominance. If we are concerned with a particular segment of the population, we can see how the probability of dominance changes if only the people in that segment or below are considered.

## 4. Empirical Results

### 4.1 Parameter Estimates

The variable EDHIGH1, describing the highest level of education achieved, was extracted from the HILDA survey for the years 2001, 2006, 2014 and 2017. It is originally a categorical variable with 10 categories; some were combined to give the 7 categories listed in Table 1. A total of $M = 10,000$ draws of population proportions $p$ were sampled using the Bayesian sampling algorithm described in Appendix S1. The posterior means (with posterior standard deviations in brackets) for the population proportions of the indigenous education distributions for the years 2001, 2006, 2014 and 2017 are reported in Table 2. They reveal positive skewed distribution for all years; there are fewer people in the higher education categories: bachelor, graduate diploma, and masters or doctorate categories and more people in the lower education categories: Year 11 or below, Year 12, and certificate. In all years, the modal category is Year 11 or below. This category is also the median category in 2001 and 2006. In 2014 and 2017, the median has moved to the Year 12 category.

### 4.2 Indigenous Level of Education

To answer the question has the level of education for the indigenous population improved over time, we examine the FSD probabilities for dominance for each pair of years, restricted FSD over the first two categories, and how the $H$ index has changed over time. The posterior probabilities for FSD are presented in Table 3. Pairwise comparisons for all 4 years are considered. The first row, $\Pr[X >_{FSD} Y]$, shows the probabilities for each of the earlier years dominating each of the later years. The probabilities for each of the later years dominating each of the earlier years are given in the second row, $\Pr[Y >_{FSD} X]$. High probabilities in this row are indicative of an improving level of education. The third row gives the probabilities for no dominance. These is some evidence of improvement, particularly from 2001 to 2017 and from 2006 to 2017, but, overall, the probabilities for no dominance are relatively high.

Improvement in terms of FSD is a relatively strict criterion. When we focus on the first two categories using the restricted FSD criterion, whose probabilities are given in the last three rows of Table 3, we find strong evidence of improvement. The probabilities for 2017 dominating both 2001 and 2006 are both greater than 0.96. To illustrate the impact of restricting dominance to fewer categories, we can plot the probability curves discussed in Section 3. Because these curves give the probability of dominance at each category, and the probability of dominance over any range of categories will be no greater than the minimum value within that range, we can use these curves to see which categories have the greatest impact on the probability of dominance or lack of it. In Figure 2, we have plotted the probability curves for 2006 FSD 2001 and 2017 FSD 2014. The probability $\Pr[2006 >_{FSD}^{rest} 2001]$ is bigger than $\Pr[2006 >_{FSD} 2001]$ because the minimum value in the probability curves occurs at the third category. For 2017 FSD 2014, the minimum value occurs at category 4. Thus, restricting the analysis to the first three categories would yield a higher probability of dominance.

The final criterion for assessing improvement in the level of education for the indigenous population is the headcount ratio. Its posterior densities for each of the years are plotted in Figure 3 and its posterior means and standard deviations are presented in the first column of Table 4. We observe little change from 2001 to 2006, but a reduction in the point estimate from 0.70 to 0.59 from 2006 to 2017, implying from 2006 onwards a greater proportion of the population was obtaining a tertiary education. Although the headcount index did not change from 2001 to 2006, there was, nevertheless, an improvement in the bottom of the distribution. From Table 2, we observe the proportion in category 1 (year 11 and less) went down, and there was a corresponding increase in category 2 (year 12).

*4.3    Inequality between Indigenous and Non-indigenous Populations*

To examine the gap between the level of education for the indigenous and non-indigenous populations and to assess whether that gap has narrowed, we again use FSD and the $H$ index, but, instead of making comparisons overtime, we compare the two populations. The FSD probabilities in Table 5 provide strong evidence that the education distribution for the non-indigenous population dominates that for the indigenous population, and there is no evidence that the gap has narrowed. Except for 2001, where the probability is 0.85, the probabilities for dominance in this direction are all greater than 0.95 and they do not decline over time. The same probabilities for restricted dominance over the first two categories are higher; they are all greater than 0.98.

When we focus on the first two columns in Table 4 that compare the indigenous and non-indigenous headcount indices, we find that the indices for both populations have declined over time, but the gap between them has not narrowed. There has been a slight increase in the difference from 0.11 in 2001 to 0.15 in 2017.

*4.4   Educational Inequality within the Indigenous Population*

In Section 4.2 we found evidence of improvement in the level of education in the indigenous population, particularly in the lower part of the distribution. Now we ask whether that improvement has resulted in greater inequality. The tools we use for this purpose are the indices $CF$ and $J$ and GLD. The GLD probabilities in the first row of Table 6 are all greater than 0.6, providing evidence that each of the later years is dominated by each of the earlier years. Inequality has increased. This is particularly apparent from the dominance probability for 2001 GLD 2017 which is greater than 0.99. Results for the $CF$ and $J$ indices confirm this outcome. Their posterior densities are plotted in Figures 4 and 5, respectively, with $CF(0.1)$ chosen to represent the $CF$ family. There is considerable overlap in these distributions, but there is nevertheless a distinct movement to the right from 2001 to 2017. Their posterior means and standard deviations, along with those for $CF(0.9)$ are reported in Table 4, columns 3, 5 and 7.

*4.5   Inequality Within the indigenous and Non-indigenous Population*

To compare the level of inequality within each of the populations, we present the $J$, $CF(0.1)$ and $CF(0.9)$ indices in Table 4, and the GLD dominance probabilities for each year in Table 7. It is evident that educational inequality is greater within the non-indigenous population than within the indigenous population. The GLD dominance probabilities for all years for indigenous equality dominating non-indigenous equality are greater than 0.99. Similar conclusions are reached by comparing the indigenous and non-indigenous columns in Table 4. Inequality is increasing in both groups, but that for the non-indigenous population always remains higher.

## 5. Conclusions

We have employed techniques for analysing ordered categorical distributions to study changes in the level of education, and inequality in the educational distribution, of indigenous Australians over the period 2001 to 2017. Comparisons of distributions over time and with those of non-indigenous Australians were made. Particularly novel was the use of posterior probabilities of first-order-stochastic dominance and generalized Lorenz dominance to compare levels and inequalities of distributions, respectively. We find strong evidence for an improvement in the lower part of the educational distribution for the indigenous population, and more qualified support for improvement in their complete distribution. This improvement did not, however, narrow the gap between the indigenous and non-indigenous distributions. The level of education in the non-indigenous distribution also improved, maintaining the gap that existed in 2001. Inequality increased in both distributions and was consistently higher in the non-indigenous distribution. The results point to a trade-off between the level of education and inequality. Having increasing inequality caused by too many individuals being left behind in the bottom part of the distribution is clearly undesirable. However, having all individuals in the same category – the requirement for zero inequality – is also undesirable and clearly not realistic. Hence, increasing inequality is not necessarily bad if it is accompanied by increases in the level of education. In terms of the government's goal of reducing the proportion of indigenous individuals in the bottom two categories to 0.3 by 2031, there is still a long way to go. Our estimates tell us that, between 2001 and 2017, the reduction was from 0.70 to 0.59.

## 6. Acknowledgements

This research was supported by RevITAlising (RITA) research grant from the University of Wollongong (IV032)

**References**

Allison, R.A., and J.E. Foster (2004), "Measuring Health Inequality using Qualitative Data", *Journal of Health Economics*, 23, 505-524.

Anderson, G., Maria G. Pittau, and R. Zelli (2020), "Measuring the Progress of Equality of Educational Opportunity in Absence of Cardinal Comparability", Metron, 78, 155-174.

Angelico, T. (2020), "The Pandemic and Educational Inequality in Australia: Timely Opportunity to Reform Education", *Journal of Higher Education Theory and Practice*, 20(14), 105-110.

Apouey, B., J. Silber, and Y. Xu (2020), "On Inequality-sensitive and Additive Achievement Measures Based on Ordinal Data", *Review of Income and Wealth*, 66 (2), DOI: 10.1111/roiw.12427.

Cowell, F.A. and E. Flachaire (2017), "Inequality with Ordinal Data", *Economica*, 84, 290-321.

Dean, J.M. (2019), *In Search of the Common Wealth: Indigenous Education Inequality in Australia*, PhD thesis, Faculty of Education, University of Canberra.

Gelman, A. J. B. Carlin, H.S. Stern, and D.B. Rubin (1995), *Bayesian Data Analysis*, New York: Chapman & Hall.

Gunawan, D., W.E. Griffiths, and D. Chotikapanich (2018), "Bayesian Inference for Health Inequality and Welfare using Qualitative Data", *Economics Letters*, 162, 76-80.

Gunawan, D., A. Panagiotelis, W.E. Griffiths and D. Chotikapanich (2020), "Bayesian Weighted Inference from Surveys", *Australian and New Zealand Journal of Statistics*, 62(1), 71-93.

Jenkins, S. (2020), "Inequality Comparisons with Ordinal Data", *Review of Income and Wealth*, DOI: 10.1111/roiw.12489.

Lander, D., D. Gunawan, W.E. Griffiths and D. Chotikapanich (2020), "Bayesian Assessment of Stochastic and Lorenz Dominance", *Canadian Journal of Economics*, 53(2), 767-799.

Perry, L. (2018), Educational Inequality in Australia, in *How unequal? Insights into Inequality*, Committee for Economic Development of Australia, pp. 56-67.

Sen, A. (1987) *Standard of Living*, Oxford: Oxford University Press.

Thomas, V., Y. Wang, and X. Fan (2001), "Measuring Education Inequality: Gini Coefficients of Education", *Research Working Paper no.2525,* World Bank, Washington D.C.

Watson, N. and Wooden, M. 2012. The HILDA Survey: A Case Study in the Design and Development of a Successful Household Panel Study, Longitudinal and Life Course Studies, vol. 3, no. 3, pp. 369–381.

Table 1: Categories for Highest Education Level Achieved

| Category | Explanation |
|---|---|
| 1 | Year 11 or below or undetermined |
| 2 | Year 12 |
| 3 | Certificate I, II, III or IV, and Certificate not defined |
| 4 | Advanced Diploma, Diploma |
| 5 | Bachelor or Honours |
| 6 | Graduate Diploma, Graduate Certificate |
| 7 | Postgraduate – Masters or Doctorate |

Table 2: Posterior Means (Standard Deviations) for the Population Proportions of the Indigenous Education Distribution

| $k$ | Category | Parameter | 2001 | 2006 | 2014 | 2017 |
|---|---|---|---|---|---|---|
| 1 | Year 11 or below | $p_1$ | 0.5876 (0.0522) | 0.5401 (0.0502) | 0.4904 (0.0381) | 0.4034 (0.0367) |
| 2 | Year 12 | $p_2$ | 0.1141 (0.0351) | 0.1599 (0.0372) | 0.1639 (0.0288) | 0.1875 (0.0290) |
| 3 | Certificate | $p_3$ | 0.1871 (0.0425) | 0.2097 (0.0413) | 0.2342 (0.0298) | 0.2748 (0.0347) |
| 4 | Diploma | $p_4$ | 0.0325 (0.0184) | 0.0170 (0.0134) | 0.0308 (0.0136) | 0.0484 (0.0158) |
| 5 | Bachelor | $p_5$ | 0.0532 (0.0268) | 0.0525 (0.0227) | 0.0557 (0.0165) | 0.0448 (0.0155) |
| 6 | Graduate Diploma | $p_6$ | 0.0182 (0.0135) | 0.0146 (0.0119) | 0.0112 (0.0084) | 0.0182 (0.0104) |
| 7 | Postgraduate | $p_7$ | 0.0073 (0.0090) | 0.0062 (0.0081) | 0.0137 (0.0084) | 0.0229 (0.0103) |

Table 3: First Order Stochastic Dominance Probabilities for Indigenous Population, 2001-2017

|  | 01 | 06 | 01 | 14 | 01 | 17 | 06 | 14 | 06 | 17 | 14 | 17 |
|---|---|---|---|---|---|---|---|---|---|---|---|---|
|  | X | Y | X | Y | X | Y | X | Y | X | Y | X | Y |
| $\Pr[X >_{FSD} Y]$ | 0.0949 | | 0.0129 | | 0.0004 | | 0.0288 | | 0.0003 | | 0.0050 | |
| $\Pr[Y >_{FSD} X]$ | 0.0896 | | 0.2571 | | 0.4961 | | 0.2861 | | 0.5541 | | 0.4010 | |
| Pr(No Dominance) | 0.8155 | | 0.7300 | | 0.5035 | | 0.6851 | | 0.4456 | | 0.5940 | |
| $\Pr[X >_{FSD}^{rest} Y]$ | 0.2471 | | 0.0543 | | 0.0011 | | 0.1286 | | 0.0046 | | 0.0349 | |
| $\Pr[Y >_{FSD}^{rest} X]$ | 0.4246 | | 0.7474 | | 0.9629 | | 0.7007 | | 0.9699 | | 0.8618 | |
| Pr(No Dominance) | 0.3283 | | 0.1983 | | 0.0360 | | 0.1707 | | 0.0255 | | 0.1033 | |

*Notes:* 01, 06, 14 and 17 refer to the years 2001, 2006, 2014 and 2017. *rest FSD* refers to restricted dominance over categories 1 and 2.

Table 4: Headcount and Inequality Measures

| Year | H | | J | | CF(0.1) | | CF(0.9) | |
|---|---|---|---|---|---|---|---|---|
|  | Ind | N-Ind | Ind | N-Ind | Ind | N-Ind | Ind | N-Ind |
| 2001 | 0.7017 | 0.5873 | 0.3730 | 0.4769 | 0.4078 | 0.5448 | 3.0436 | 3.7910 |
|  | (0.0511) | (0.0084) | (0.0380) | (0.0048) | (0.0446) | (0.0068) | (0.2578) | (0.0305) |
| 2006 | 0.7000 | 0.5446 | 0.4040 | 0.5030 | 0.4422 | 0.5832 | 3.2237 | 3.9526 |
|  | (0.0455) | (0.0089) | (0.0324) | (0.0041) | (0.0390) | (0.0064) | (0.2064) | (0.0259) |
| 2014 | 0.6543 | 0.4714 | 0.4366 | 0.5356 | 0.4835 | 0.6378 | 3.4413 | 4.1421 |
|  | (0.0339) | (0.0073) | (0.0221) | (0.0025) | (0.0286) | (0.0046) | (0.1395) | (0.0152) |
| 2017 | 0.5909 | 0.4456 | 0.4806 | 0.5454 | 0.5444 | 0.6569 | 3.7032 | 4.2028 |
|  | (0.0365) | (0.0079) | (0.0158) | (0.0020) | (0.0238) | (0.0041) | (0.1006) | (0.0129) |

*Note:* Ind refers to the indigenous population; N-ind refers to the non-indigenous population.

Table 5: First Order Stochastic Dominance Probabilities for Indigenous ($X$) versus Non-indigenous ($Y$) Population

|  | 2001 | 2006 | 2014 | 2017 |
|---|---|---|---|---|
| $\Pr[X >_{FSD} Y]$ | 0.0000 | 0.0000 | 0.0000 | 0.0000 |
| $\Pr[Y >_{FSD} X]$ | 0.8477 | 0.9512 | 0.9831 | 0.9655 |
| Pr(No Dominance) | 0.1523 | 0.0488 | 0.0169 | 0.0345 |
| $\Pr[X >_{FSD}^{rest} Y]$ | 0.0006 | 0.0001 | 0.0000 | 0.0000 |
| $\Pr[Y >_{FSD}^{rest} X]$ | 0.9856 | 0.9976 | 1.0000 | 0.9998 |
| Pr(No Dominance) | 0.0138 | 0.0023 | 0.0000 | 0.0002 |

*Note*: *rest FSD* refers to restricted first order stochastic dominance over categories 1 and 2.

Table 6: Generalized Lorenz Dominance Probabilities for Indigenous Population, 2001-2017

|  | 01 X | 06 Y | 01 X | 14 Y | 01 X | 17 Y | 06 X | 14 Y | 06 X | 17 Y | 14 X | 17 Y |
|---|---|---|---|---|---|---|---|---|---|---|---|---|
| $\Pr[X >_{GLD} Y]$ | 0.6451 | | 0.8964 | | 0.9933 | | 0.7319 | | 0.9761 | | 0.8822 | |
| $\Pr[Y >_{GLD} X]$ | 0.2358 | | 0.0466 | | 0.0007 | | 0.1510 | | 0.0037 | | 0.0323 | |
| Pr(No Dominance) | 0.1191 | | 0.0570 | | 0.0060 | | 0.1171 | | 0.0202 | | 0.0855 | |

Table 7: Generalized Lorenz Dominance Probabilities for Indigenous ($X$) versus Non-indigenous ($Y$) Population

|  | 2001 | 2006 | 2014 | 2017 |
|---|---|---|---|---|
| $\Pr[X >_{GLD} Y]$ | 0.9976 | 0.9984 | 1.0000 | 0.9996 |
| $\Pr[Y >_{GLD} X]$ | 0.0000 | 0.0000 | 0.0000 | 0.0000 |
| Pr(No Dominance) | 0.0024 | 0.0016 | 0.0000 | 0.0004 |

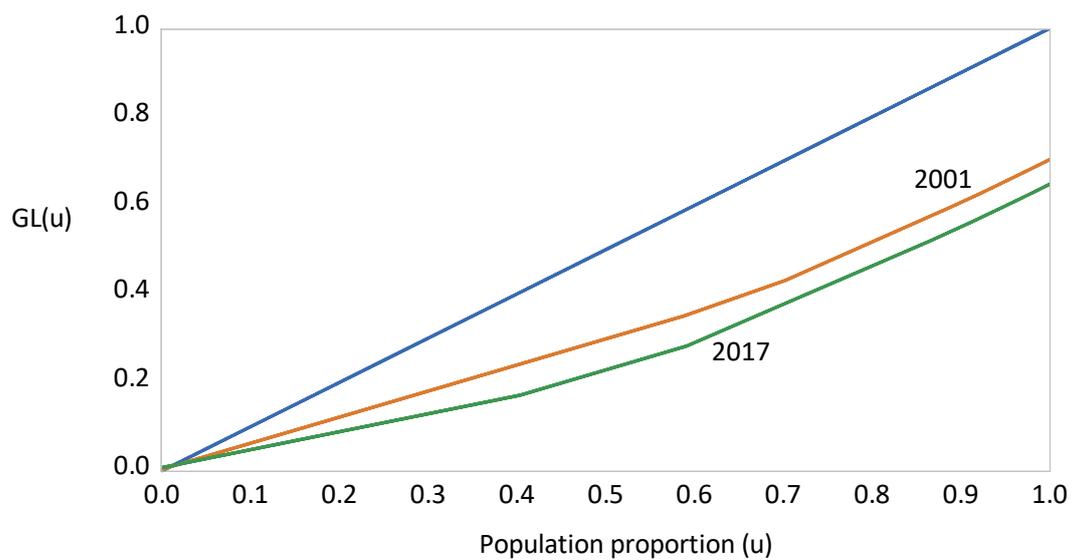

**Figure 1** Indigenous population generalized Lorenz curves for 2001 and 2017

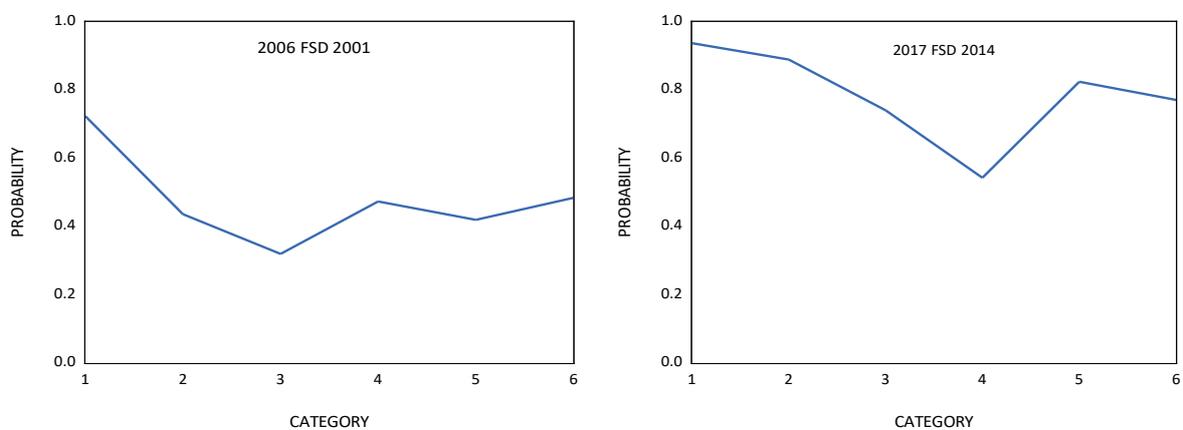

**Figure 2** FSD Probability Curves

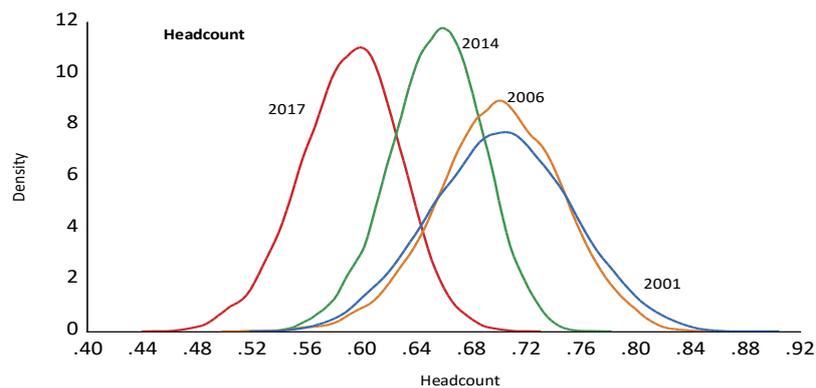

**Figure 3 Posterior Densities for the Headcount for the Indigenous Population**

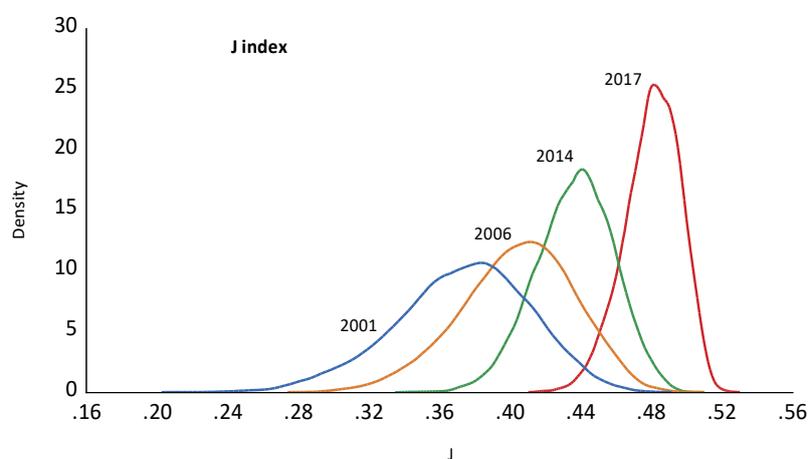

**Figure 4 Posterior Densities for the $J$ Index for the Indigenous Population**

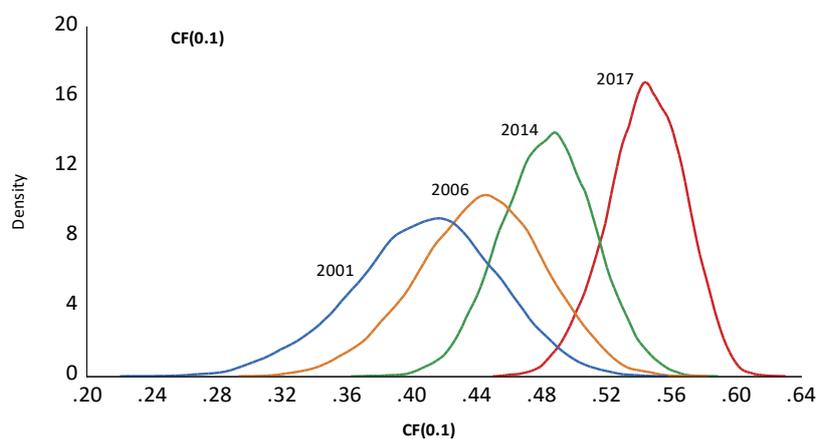

**Figure 5 Posterior Densities for the $CF(0.1)$ Index for the Indigenous Population**